# Blockchain Technology: Methodology, Application and Security Issues

**AKM Bahalul   Haque, Mahbubur Rahman**

North South University, Dhaka Bangladesh

**Summary**

Blockchain technology is an interlinked systematic chain of blocks that contains transaction history and other user data. It works under the principle of decentralized distributed digital ledger. This technology enables cryptographically secure and anonymous financial transactions among the user nodes of the network enabling the transactions to be validated and approved by all the users in a transparent environment. It is a revolutionary technology that earned its emerging popularity through the usage of digital cryptocurrencies.  Even though Blockchain holds a promising scope of development in the online transaction system, it is prone to several security and vulnerability issues. In this paper, blockchain methodology, its applications, and security issues are discussed which might shed some light on blockchain enthusiasts and researchers.

## 1. Introduction

Blockchain technology is a peer to peer architecture network. It is decentralized and comprised of a series of blocks known, hence it is called blockchain. After the initial concept derived and implemented by Satoshi Nakamoto in bitcoin, Blockchain has become a topic of interest among the researchers [1][2]. Moreover, its characteristics have expanded the applicability to a greater extent.

It is also referred to as distributed ledger technology, which preserves the calculation of all the nodes in each of them. Since the ledger is shared; reliability is not a concern in the network. Moreover the blocks include hash code, which is a unique and unchangeable value derived using complex mathematical hash function. For this reason, immutability is ensured [3].Among other characteristics, transparency is ensured by the reasons mentioned above. As the transaction does not happen in traditional way as in with individual real user id and address, there are several scopes to make both the sender and the receiver anonymous. The absence of central authority makes the whole system autonomous to some extent [4]. These reasons have made the concept of blockchain as an emerging technology to be implemented in various fields.

## 2. Various Types OF Blockchain

Blockchain is the foundation of the digital cryptocurrency, Bitcoin. It has raised its sheer importance in the digital world by holding its critical character traits of decentralization, immutability, anonymity, and suitability for the e-money transaction process. There are primarily four types of blockchains to be considered.

### 2.1 Public Blockchains

A public blockchain is an open-source, decentralized blockchain with no restriction of users that can participate in the network. No individual entity has control over the network instead anyone can join the network and read/write/audit the blockchain with no order for processing the transactions [5]. Public blockchains, by its design, are the best for protecting user anonymity. Since this type of blockchain is publicly accessible to all users, the decisions here are made by several consensus algorithms such as Proof of Work (POW), Proof of Stake (POS), and many more [6]. Moreover, the public blockchain platform maintains an incentive mechanism predefined in the protocol through some gaming theory, that is, the participants in the network are economically rewarded for maintaining the best of behaviors and honesty in the system [7]. Public Blockchain platforms include Bitcoin, Ethereum, Litecoin, etc.

### 2.2 Private Blockchains

Private blockchain restricts the users who can participate and make a transaction in the network as shown in Fig. 3. A group of individuals or organizations that are permitted to enter the network holds the control of the blockchain network. Thus a private blockchain from the very beginning has user identity to some extent for determining their respective tasks in the networks and their controlling access such as reading/writing/auditing of specific information in the blockchain. The design of the private blockchain, in contrast to the public blockchain, is more centralized, so the decisions are made by an in-charge who assigns several rights to the participants in the network [6]. However, the centralized architecture of a private blockchain makes it more prone to security breaches. The private blockchain is used by organizations or enterprises that require scalability, data protection privacy, and regulatory rules for state compliance [7]. So, selective participants who are pre-defined with specific criteria in the network have access to the blocks of information for internally verifying and validating the transactions. Private blockchain platforms include Hyperledger, Hashgraph, Corda, and many others.

## 2.3 Consortium/Federated Blockchains

A consortium blockchain is a partially decentralized blockchain, which means it lies in between a public and a private blockchain. Consortium blockchain partially exhibits the properties of both public and private blockchains, but it retains most of the characteristics of a private blockchain. Unlike a private blockchain, the network of a Consortium blockchain is operated by a group of entities [4]. In contrast to a public blockchain, a Consortium blockchain does not allow anybody to enter the network instead it requires the permission of the network admins to grant access to someone. This type of blockchain is often referred to as "semi-private" blockchain since the authority of the network is given in advance to a selectively predefined nodes based on several consensus algorithms. Generally, it is used in business organizations having several business partners. Examples of consortium platforms include R3, Corda, etc.

## 2.4 Hybrid Blockchains

A hybrid blockchain is a combination of both the public blockchain and private blockchain. It combines the advantages characteristics of each blockchain, respectively, that is, a hybrid blockchain inhibits the privacy benefits of private blockchain and transparency benefits of a public blockchain according to necessity. The patented Interchain ability gave rise to the hybrid nature of the blockchain enabling the hybrid blockchain to have multiple chain networks of blockchains [5]. The hybrid blockchain is not entirely open to everyone; certain restrictions are posted while allowing participants to enter the hybrid blockchain network. This hybrid characteristic of this platform gives organizations the properly controlled access of their data in the network thus making the system more flexible and secure without compromising the privacy. Even though the hybrid blockchain is controlled by a group of individuals, the transactions made are kept private and yet can be verified whenever needed. This upholds the immutability of the transactions [8]. Hybrid blockchain platforms include Dragonchain.

## 3. Methodology

Blockchain technology works by creating an environment that is secure and transparent for the financial transactions of virtual values such as Bitcoin. Hash codes of each block keep records safe in the blockchain. This is mainly because irrespective of the size of the information or document, the mathematical hash function provides a hash code of the same length for each block. So, attempting to change a block of information would generate a completely new hash value [9].

A network that is open to everyone and concurrently maintains user's anonymity undoubtedly raises trust issues regarding the participants. So, to build the trust the participants need to go through several consensus algorithms such as Proof of Work and Proof of Stake.

The digital cryptocurrency Bitcoin uses the first-ever blockchain technology [10]. It is a digital store of value that enables peer to peer transactions over the internet without the intervention of a third party. The blockchain network is a decentralized structure that consists of scattered nodes (computers) that inspect and validate the authenticity of any new transactions that attempt to take place. This combine agreement is done through several consensus models by the process of mining. The process of mining demonstrates that each node trying to add a new transaction has gone through and solved the complex computational puzzle through extensive work and deserves to get a reward in return for their service. For the validation of a transaction, the network must confirm the following conditions:

The sender account holds sufficient Bitcoin balance that it intends to transfer. The amount intended to transfer has not already been sent to some other recipient.

Once a transaction has been validated and agreed upon by all the nodes, it then gets added to the digital ledger and protected using cryptography that uses a public key that is accessible to all the other nodes and a private key that must be kept secret [11]. The transaction process in Blockchain network is shown in Fig. 1 [12].

To maintain the transactions using digital currency in the blockchain network, we need to have an understanding of the digital wallet which is used to store, send, and receive digital currency. A digital wallet or a cryptocurrency wallet is a string of letters and numbers forming a public address associated with each block in the blockchain. This public address is used whenever a transaction takes place; that is, the Bitcoin currency is assigned to the public address of the specific wallet. However, to prove the ownership of the public address there is a private key associated with the wallet that serves as the user's digital signature that is used to confirm the processing of any transaction. The user's public key is the shortened version of his private key generated through complex and advanced mathematical algorithms [13].

For example, let us consider someone is trying to send you some digital currency such as Bitcoin, as the transaction is being processed, the private key in your wallet should match the crucial public address of your wallet that the currency has been assigned to. If both these keys match, then the digital currency amount is transferred to the public address of your wallet.

## 3.1 Block Structure

A block contains several parts [14] as shown in Fig. 2 [4].

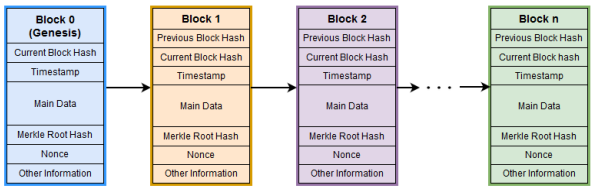

Fig. 1  A sequnce of blockchain showing block structure

Main Data: Blocks will contain transaction data. This transaction data depends on the usage factor of blockchain, that is, the relevant services for which the blockchain is implemented. For financial institutions like banks, financial transaction data will be stored.

Timestamp: The timestamp will also exist in the blocks. Here, the timestamp refers to the date and time when a particular block is generated.

Hash: The hash corresponding to each block is a unique identifier that is generated using a cryptographic hash algorithm such as SHA-256. Hash of the current block and hash of the previous block will be stored in the block. Hashes make the blocks immutable. Hashes are generated using the Merkle tree function. It is stored in the header of the block.

Merkle tree root hashIt consists of all the hash values relating to every transaction that took place in a block and performs a mathematical hash calculation generating a 64-character code [15]. The hash of the Merkle tree root of all the transactions in the block is stored for effective processing and easier verifying of data within a short time.

NonceA nonce is a randomly generated 4-byte number that can be used once in a cryptographic transaction process. During the mining process in a Proof-of-Work algorithm, the nonce is used as a counter that the miners are trying to solve in order to generate a new block. The aim is to calculate a hash value less than a given target value, which depends on the difficulty of the complex mathematical problem.

Block Properties: Each block inside blockchain mainly consists of three parts, such as Hash of the previous block, Data, hash of current block as shown in Fig.2. Data on the block can be anything. It can be transaction records, medical records, insurance records, law records, property ownership records, etc. There are mainly two types of Blockchain. One is Private Blockchain; another is public blockchain. There is one more, which is a mix of Private and Public blockchain called Hybrid blockchain shown in Fig.3.

Each block is connected to previous block using the Hash value shown in Fig.4. Changing the value of a single data in a block will result in a change in the HASH value of that block. If a block has been changed or not, it can be easily verified by checking the hash of that block. Hash of a block is like a fingerprint. Every block has a unique hash number, just like a human fingerprint.

As each block contains the hash of previous block tampering with one block will change its hash, and it will be disconnected from all the forwarded blocks from its position as shown in Fig.5. Once a new block is created it is announced in the network. All the nodes then verify the block and add it to their list of blocks as shown in Fig.6. is a fraudulent block can also be sent out in the network. To check if a block is real or not, its hash value is checked. The hash value of a block needs to be a specific range so that others can know computational work has been done on this particular block. There is a delay regarding the addition of blocks. So that fake blocks cannot be added instantly one after another. Bitcoin has about 10 minutes delay between each block creation.

However, if some individual or mining pool has more than 51% computational power of that network, and they keep sending fraudulent blocks and also calculate the subsequent hashes of the following blocks. This is a significant issue with PoW consensus in the blockchain. That is why some model will use proof of stake instead of proof of work.

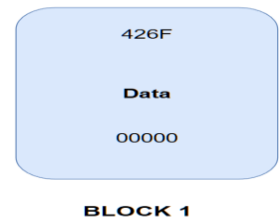

Fig. 2  Contents of Block

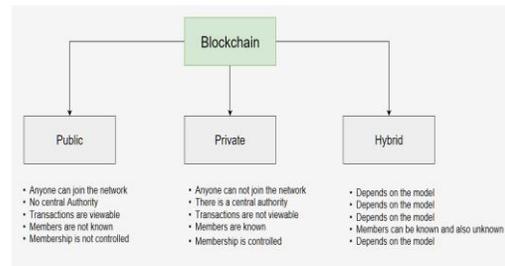

Fig. 3  Different types of blockchain and their attributes

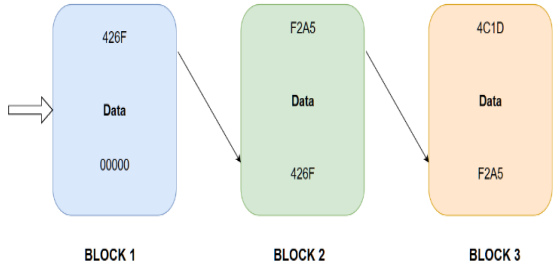

Fig. 4 Chain of Blocks Using Hash Values

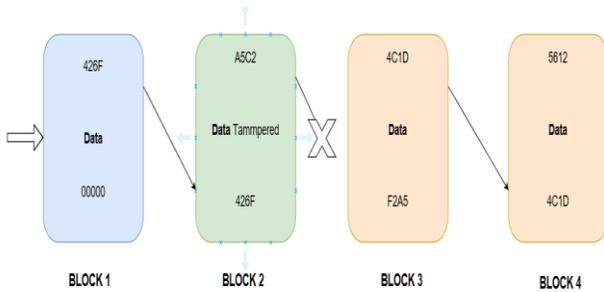

Fig. 5 Tampered Block disconnected from all forwarding block

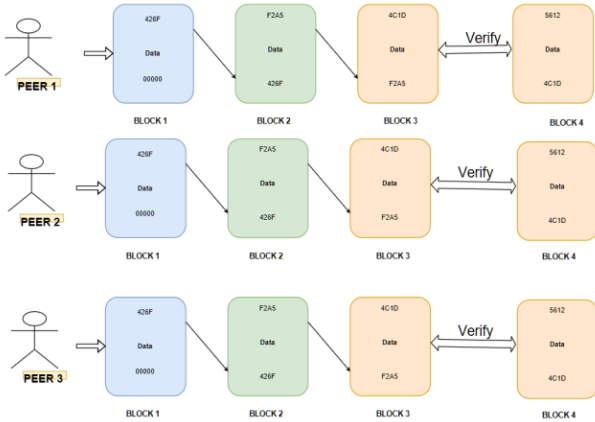

Fig. 6 Block Verification

Hash Function: A hash function takes an input and returns a fixed-length output, e.g. SHA1 [16]. The output of the hash function is different for different messages and the same for the same input. A hash function has some internal states. Based on the message it receives, it will change those internal states. Through permutations and combinations, the internal states will change in such a way it is quite impossible to guess the input message from the hash output. This means knowing the output we cannot know or guess the output. Hash of a block in blockchain technology takes a large quantity of computer power.

Changing the input slightly changes the hash output wildly. There is no rule on how these changes occur and appear to be random. Nevertheless, it is nothing but random. Still, no one yet has come up with a solution to crack the rules behind how changes if the input is changed. The hash algorithm is made in such a way that it is not reversible.

From Fig. 7, we can see that when if block A's special number incremented by one, the hash output of block 'A' changes dramatically. There is no correlation between the one increment in the unique number and the hash output. Though the 20-bit hash is not the actual output of a hash algorithm it is helpful for understanding. The particular number as shown in Fig. 7, is known as the nonce of a block.

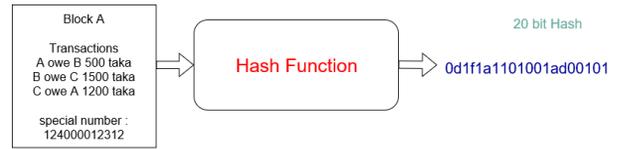

Note that the hash of this block is not the actual hash

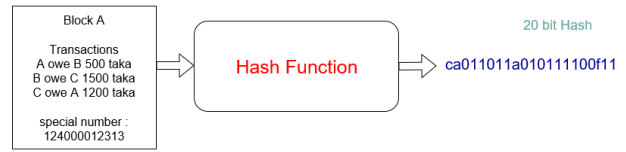

Same Block with different special number

Fig. 7 An imaginary 20-bit hash algorithm

Consensus Algorithm. The most popular consensus algorithm is Proof of Work (PoW). Nevertheless, PoW requires enormous computational power and consumes energy in the process of solution searching.

Proof of Work When a block is added to the network, Blockchain algorithm checks if that intended block is fraudulent or not. We know in DLT network trust inbuilt on the system, not on any user. A miner might throw a fraudulent block in the blockchain, how will blockchain network know that this new block in actually fraudulent? The answer is proof of work algorithm.

Proof of Work checks if the Hash value of the block is in a specified range. If the value is in range then the block is accepted else rejected. Miners compete with each other in the PoW concept to mine blocks, which means to find a hash value of a block in a specified range.

PoW is a race between the miners to find the solution to a mathematical puzzle. Whoever gets to the answer of that puzzle, in other words, finds the hash value with a nonce that falls in a specific range gets the reward for mining that block. To add a malicious block in a PoW network, one needs to have at least 51% computational power than the whole network. Otherwise, one trying to add maliciously would not be successful.

Proof of Stake

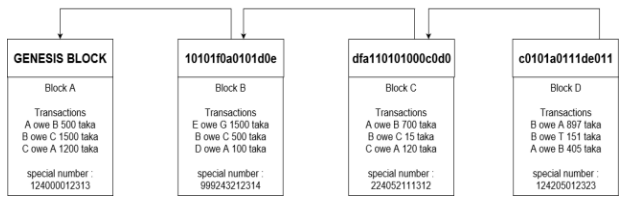

Fig. 8  Blocks in Blockchain

In Proof of Stake the job of the creation of a new block is given to a node has the most stake. A node is less likely to put out a fraudulent block if the stake of that node is high. They simply would not risk that. Proof of Stake is much faster than Proof of Work, and it consumes significantly less power than proof of Work. For these reasons, its why many people use proof of stake in their model. [17].

In PoW, the mining job is a race between the miners. The race is to find a solution to the cryptographic hash function. In PoS block creator is chosen by the PoS algorithm based on stakes. Unlike PoW, there is no reward for creating a block in PoS. The block owner instead gets transaction fees. Similar to PoW for block generation, PoS also needs to generate a hash value. Generation of this hash value, just like PoW, is random. Unlike PoW, the search for this hash value is not over any real number domain but the specified range. So, the search space for hash value in PoS is limited thus making PoS faster and more energy-efficient than PoW. To add a malicious block in PoS attacker would need to have at least 51% currency of that network. Uses significantly less power than a proof of work as block provides following opportunity-

Adding the malicious block requires the hacker to have more than 50% of the currency of that network, which is quite unlikely [18].

# 4. Application OF Blockchain

Blockchain application can be applied to many sectors in Bangladesh. It either uplifts the existing process or creates new technologies. It will change our lives and protect us from fraud, thief, or any crime. Blockchain provides a secure way of sending digital assent without even knowing third parties. In many sectors, we can use blockchain. Some of the sectors are:

## 4.1 Healthcare

Security of Personal information like health data is very important for everyone. These health related information are very valuable as pharmacy companies thrives on these data. Most of the time these data are kept on a hospital server and not in a secured environment too. To prevent health data from falling on the wrong hands and to prevent misuse of these data public blockchain can be used where health data can on be seen by the doctors if the owner of the data i.e. the patient permits it [19]. Building a system like this will make sure proper security for these data.

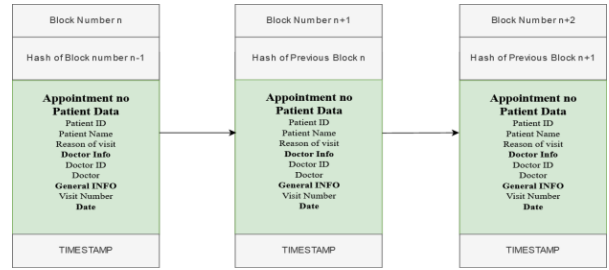

Fig. 9  Structure of medical blocks

## 4.2 Equity Crowdfunding

Getting money from crowd or supporters of a company/product in exchange of equity/shares in that company known as equity crowdfunding. As people from different background participate in these crowdfunding and they follow different rules and regulations, it becomes very tough to maintain policies. Also not all people will trust the party that is handling all the transactions thus affecting the total amount of money generated. To maintain a fare ground among fundraisers and investors blockchain technology can be used. Blockchain will remove differences due to regulatory laws and bring all the parties together in a systematic way [20].

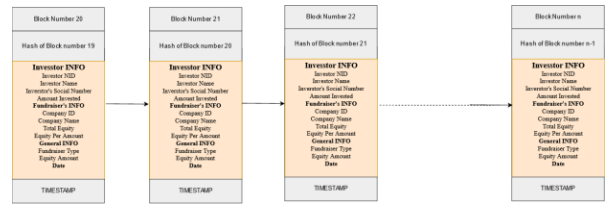

Fig. 10  Structure of crowdfunding blocks

## 4.3 Banking

It needs to use blockchain in the tax sector. Blockchain can make a significant contribution in the tax sector. Our current tax collected system is not so trusted. Blockchain can change the whole system, the way our tax is collected. As blockchain provides accurate information, so if we use this in the tax sector then it will be beneficial for all the people. Sometimes people are argued about paying taxes. They denied paying taxes and claim false information. Blockchain is an immutable system. It gives the real value of the system. When someone lying about paying tax, blockchain can immediately give the correct information about tax. As this is immutable people could change the information if they want. It is impossible to change the complete information, so when people lying about paying the tax they will be caught immediately. Blockchain can detect fraud easily. It can reduce VAT fraud [20]. Now, in the present system, tax authorities have to take information from taxpayers one by one. It is not only challenging but also not safe. It could not give real data. Nevertheless, if blockchain will use in tax system then it is not necessary to collect information from the taxpayer. Authorities can see the result in computer, and that result will be real and genuine. Security of money is important for everyone. So people use bank to keep our cash safe. But even banks cannot provide absolute security. But now a day's most of our money is not real money but rather digital form of money. Digital things must be kept on a server somewhere operated by software. Just like any other software banks get hacked too. In a blink of an eye money can vanish from bank. Blockchain is used to prevent hackers from getting money from bank. A banking system totally based on Blockchain is very hard to hack and steal money. Banking based on Blockchain makes sure that even if a server gets hacked hackers cannot steal the right away because they would also have to hack all the other servers that contains the same information as the hacked server.

### 4.4 Smart Power Grid

Blockchain is also used in power grid to supply electrical power to customer. This eliminates fraud. Everyone power consumption will be recorded in the Blockchain ledger. Since everyone has the same ledger no one just claim they used less power than they actually have. Authority also cannot overcharge a customer.

### 4.5 Smart Delivery System

Making sure we get the right product we ordered online can be a tough job for distribution companies. Delivery relying on third parties can be easily hacked, single point of failure. Once compromised attacker can procure item that was intended for others. Blockchain can remedy this using smart [21]. But using Blockchain for every single item is not practical. There is no need to create a Blockchain system for grocery delivery instead we need one for very valuable items like gold, statues, documents etc. Smart contract based Blockchain can make sure that the right person gets the item. Secret code can be shared between seller and buyer. Once item reaches its destination previously shared secret code will complete the contract. This way both buyer and seller can make sure that the transaction was successfully completed.

## 5. Security Issues and Challenges in Blockchain Technology

Despite cryptographic hash protection and immutability, blockchain suffers from vulnerabilities and challenges. Vulnerabilities lie within the feature itself. Some of the security issues are described below:

### 5.1 51% Attack

This attack happens due to the consensus algorithm. The consensus algorithm is responsible for selecting the miners to solve the mathematical problem. After solving the mathematical problem, the block is added to the chain, and it is broadcast all over the network. Bitcoin and other cryptocurrencies need to do this proof of work. The ledger is made public. To solve the mathematical problem, it takes a considerable amount of computing power, and the average time is 10 minutes.

Owning 50 percent of the total computing power in the network makes that specific miner or the group to control the bitcoin network. This even interferes with the other miners. Other miners might not be able to mine new blocks. So, several illegitimate blocks can be approved, which will cause a double-spending attack [22]. This is described below:

- Moreover, previous blocks can also be changed as the computing power owned by the user/user group is fairly a lot. Though it is not possible to edit all the earlier blocks and considerably earlier historical blocks as it would take a lot of computing power, time, and complex calculations to do it. A 51% attack on blockchain happens in the following ways [23].
- Whenever a new transaction is created, it is transferred to the network.
- Miners receive it and try to solve the mathematical problem, which is called hashing.
- Now the malicious miner can solve the mathematical problem and create an offspring block without broadcasting the block to the network.
- The attackers do not stop working on his malicious block but do not broadcast it to others. This results in two separate chains; one is authentic, and another one is an illegitimate chain.
- The attacker now spends all his coin in the authentic block. In the meantime, the illegitimate block starts receiving blocks and solves those due to owning more than 50% of the computational power.
- The illegitimate chain slowly gets longer as the attacker will add more blocks in its chain. The attacker will broadcast this chain to the network.

As the chain broadcasted by the attacker is longer so it will be accepted by the other nodes also as per the rule. Money spent on the authentic blockchain will be reversed as it is no longer the accepted version of the chain.

Table 1: Some of the attacks and their details.

| No. | Name of the Organization/Groups | Details of 51% attack |
|---|---|---|
| 1 | Ghash.io | Exceeded 50% during 2014. Later nodes voluntarily dropped to preserve the integrity |
| 2 | Krypton and Shift | Etherium based blockchain was attacked , August 2016 |
| 3 | Bitcoin Gold | Attacked during August 2018, Stole 18 Million dollar worth of Bitcoin Gold |

## 5.2 Unlawful incidents

Table 2: Amount of Each Category Available in SilkRoad

| Number | category | Items | Percentage(%) |
|---|---|---|---|
| 1 | Weed | 3338 | 13.7 |
| 2 | Drugs | 2194 | 9.0 |
| 3 | Prescriptions | 1784 | 7.3 |
| 4 | Benzos | 1193 | 4.9 |
| 5 | Books | 955 | 3.9 |
| 6 | Cannabis | 877 | 3.6 |
| 7 | Hash | 820 | 3.4 |
| 8 | Cocaine | 630 | 2.6 |
| 9 | Pills | 473 | 1.9 |
| 10 | Blotter(LSD) | 440 | 1.8 |

Bitcoin is anonymous. The transaction is not related to the user's authentic identity and hence is a very daunting task and tends to be impossible to find out the identity. To buy something online using online, users use third party websites. These third-party websites can be used to buy illegal materials such as weed, drugs, prescriptions, etc. A statistic is given below in tabular form about the items found in Silk Road, a popular black-market [24].

Categories mentioned in table 2 are easily found in the online marketplace Silk Road. Anyone can buy items with bitcoin. As mentioned earlier bitcoin is anonymous and cannot be tracked. So, the identity of people who are buying it is very much challenging to find out. That is why illegal trading can happen and tracking the transaction is not possible.

Wannacry ransomware attacks harmed many people in May 2017. Due to this attack the files were encrypted, and those files were not restored till a ransom has been paid [25]. The ransom was BTC 0.17 that is 300 dollar. Almost 230000 people were infected.  Another ransomware attack CTB-Locker attacked in July 2014. It was spread through mail attachments and the files were encrypted till the ransom was paid. The ransom was transferred using the bitcoin and the transaction could not be tracked.

Money laundering is a serious financial crime. Bitcoin can be stored in Darkwallet, and transfer from Darkwallet is completely anonymous as the real coins are mixed chaffed coins. For this reason, a large amount of fund transfer is possible from one party to another through bitcoin [26]. Due to the existence of blockchain illegitimate digital money transfer is possible which causes money laundering problem

## 5.3 Inefficient Transaction time

Table 3: The transaction speed of various Cryptocurrencies

| Name of the Cryptocurrency | Transaction Per Second |
|---|---|
| Bitcoin | 7 |
| Ethereum | 15 |
| EOS | 50 |
| Litecoin | 56 |
| Cardano | 250 |
| Stellar | 1000 |
| IOTA | 1500 |
| Tron | 2000 |
| Qtum | 10000 |
| Ripple | 50000 |

Blockchain uses a distributed ledger system, and it needs proof of work to execute a transfer. For this reason, a mathematical problem has to be solved which takes 10 minutes in case of bitcoin. Bitcoin can transfer only seven transactions per second (TPS). Though other cryptocurrencies can handle more than that, still the transaction time per second is somewhat limited by blockchain. After 2010 study shows that cryptocurrencies can handle a lot more TPS. A chart is given below for understanding - [27]

Another necessary calculation is transaction confirmation time. Transaction time varies from currency to currency same as the transaction time. Though there is no formal study for both of those but an overview is shown in a tabular form below:

Table 4: Transaction confirmation time for various Cryptocurrencies

| Name of the Cryptocurrency | Transaction Confirmation Time |
|---|---|
| Bitcoin | 78 minutes |
| Ripple | 4 seconds |
| Bitcoin Cash | 60 minutes |
| Litecoin | 30 minutes |
| Ethereum | 6 minutes |
| EOS | 1.5 seconds |
| Tron | 5 minutes |
| Cardano | 5 minutes |
| IOTA | 3 minutes |
| Stellar | 5 seconds |

Due to time variances, there exists a limit of using cryptocurrencies for day to day general transfer [28].

## 5.4 Cryptographic Key

Another vulnerability exists in the cryptographic key. Blockchain uses two keys- public and private keys. Private Key encrypts the data and executes the transfer. Blockchain has no central authority that is why if the private key is lost, there is no way to retrieve the key. Moreover, if the key falls into the wrong hands, there is a high chance of the node to be compromised.

## 5.5 Distributed Ledger Vulnerability

Blockchain users, distributed ledger. Each node in the network has a copy of the total transaction. If one node is vulnerable and is compromised, the transaction history will be at the hand of the attacker. This is a severe privacy and security issue. In the case of financial transactions, the financial transfer log will be compromised.

## 5.6 Forks in Blockchain

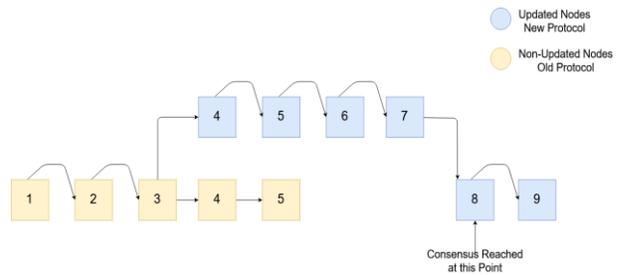

Fig. 11  Soft Fork

Like any other software Blockchain also follow some rules. These rules are called 'Protocols.' The software needs to update over time due to various reasons, e.g. bug fixes, performance enhancement, increase throughput, etc. These updates are known as Forks in the blockchain world. There are two types of forks.
- Hard Forks

Soft Forks

Hard Forks happens when the new updated Protocols are not compatible with the old Protocols. It can also happen when consensus mechanism fails [29]. When a new block is pushed by the new protocol, nodes following the old protocol will reject the node. As more and more blocks are pushed by nodes following new protocol, nodes following the old protocol can reach an agreement with them and start to follow the new protocol as shown in Fig. 10.

However, if the nodes following the old protocol do not agree with the new protocol this will permanently divide the blockchain into two chains of block each with its protocol as shown in Fig. 11.

Soft forks occur when a new protocol is compatible with the old protocol. In soft forks nodes following old protocol eventually update them with the new protocol as chain of blocks with new protocol gets bigger as shown in Fig. 9.

One disadvantage of a hard fork is that miners stop receiving fees for putting a transaction on a block. This turns out to be an advantage for a customer wanting a transaction to go through. The only way to increase transaction size on a blockchain is to do a hard fork. As transaction per block increases transaction per second also increases. This means customer has to wait for lesser amount of time. While most of forks are planned there have been instances where unplanned occurred in the blockchain world. An unplanned fork happens due to poor design of the consensus mechanism [30].

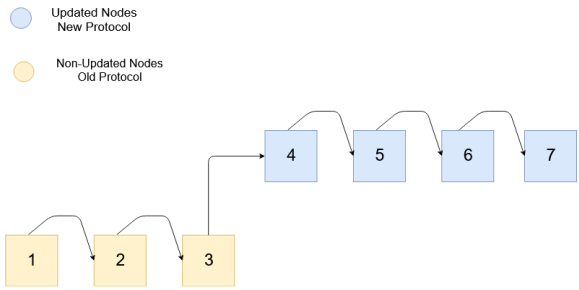

Fig. 12  Hard Fork (old nodes upgrading to new Protocol)

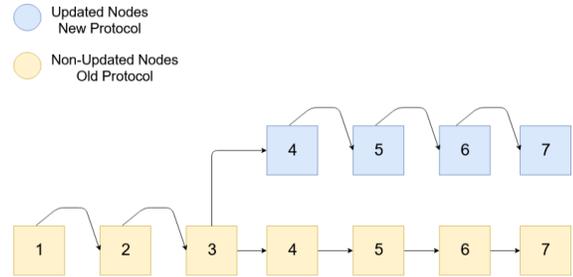

Fig. 13  Hard Fork (Dividing single Blockchain into two Chains)

**Poor Design of Consensus Mechanism**

Transaction rate and Transaction confirmation time depend on the consensus mechanism. Poor consensus design will increase transaction time and transaction confirmation time and likely to produce many problems shown in Fig. 12. Real-world problems that require instant transaction requires a good consensus that facilitates fast transaction. It's crucial for application that uses features such as sending money, file Transfer to do transaction as fast possible. A good consensus mechanism is a pre-requisite for the inclusion blockchain technology in these types of applications. The consensus mechanism can face problems like not reaching a consensus due to various things like network error leads to transaction on hold [31]. Poor consensus also leads to forking of blockchain. These type of forks are unwanted and unplanned.

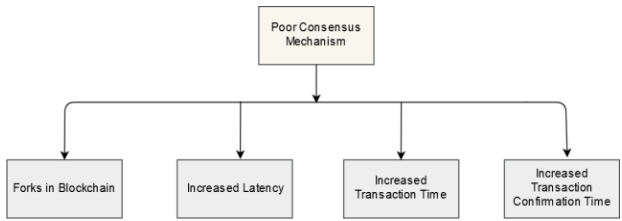

Fig. 14  Blocks in blockchain

# 6. Conclusion

The future of Blockchain predictively holds some significant advancements in technology. As the future scope, the foremost priority is to handle the several security issues that arise from different types of blockchain network such as private blockchain network which is often implemented by a business organization and big enterprises. The concept of private blockchain makes the network centralized thus making the network vulnerable to cyberattacks.

Moreover, consensus algorithms such as PoW implemented in blockchain has several drawbacks. It requires enormous amount of energy for the computation of hash. So trying to develop improved consensus algorithm would result in a cost-effective and more efficient blockchain network [32].